\documentstyle[aps,preprint]{revtex}
\begin{document}
\title{Superconductivity in the cuprates}
\author{C.D. Batista and A.A. Aligia}
\address{Centro At\'{o}mico Bariloche and Instituto Balseiro \\
Comisi\'on Nacional de Energ\'{\i}a At\'omica\\
8400 Bariloche, Argentina.}
\maketitle
\begin{abstract}
We evaluate numerically several superconducting correlation
functions in a generalized $t-J$ model derived for hole-doped
CuO$_2$ planes.  The model includes a three-site term $t''$
similar to that obtained in the large $U$ limit of the Hubbard
model but of opposite sign for realistic O-O hopping.  For
realistic parameters we obtain strong evidence of superconductivity
of predominantly $d_{x^2-y^2}$ character.  The ground state has a large
overlap with a very simple resonating-valence-bond wave function
with off-diagonal long-range order. This function reproduces the main
features of the magnetic and superconducting correlation functions.
\end{abstract}
\newpage
\par In spite of a considerable effort, no indisputable evidence
of superconductivity has been found in strongly correlated models
for the cuprates for realistic parameters.  The numerical methods
are so far the most reliable to treat strong correlations, and
clear indications of $d_{x^2-y^2}$ superconductivity have been
found in the $t-J$ model for a $4 \times 4$ cluster, but only for
values of $J$ which are an order of magnitude larger than the
realistic values and for too large doping $x = 0.5$ \cite{dag}.
Variational Monte Carlo results in the large $U$ limit of the
Hubbard model suggest that it might be necessary to solve larger
clusters to obtain superconductivity for realistic parameters, and
favor mixed $s-d$ superconductivity \cite{chen}.  On the other
hand, Monte Carlo results in the three-band Hubbard model
\cite{eme,lit} with O-O hopping $t_{pp} = 0$ favor $s$-wave
superconductivity, although the dependence of the superconducting
correlation functions with distance has not been investigated \cite{dopf}.
In this Letter we present numerical
evidence of superconductivity and at the same time a
resonating-valence-bond (RVB) state
\cite{and} in a realistic model.
\par After the original derivation of the $t-J$ model \cite{zr}, it
became clear that this model should be supplemented by other terms
to accurately represent the physics of the three-band Hubbard model
\cite{good,hyb,ding,bat,ali}, and experimental data
\cite{ali,chi,japo}.  Including the most important terms, the
Hamiltonian can be written in the form \cite{bat}:
\begin{eqnarray}
H & = & t \sum_{i \delta \sigma} c^{\dag}_{i+ \delta \sigma} c_{i
\sigma} + t' \sum_{i \gamma \sigma} c^{\dag}_{i + \gamma \sigma}
c_{i \sigma}\nonumber \\
& + & t'' \sum_{i \delta \neq \delta ' \sigma} c^{\dag}_{i + \delta
' \sigma} c_{i \delta \sigma} ({1 \over 2} - 2{\bf S}_i \cdot  {\bf S}_{i +
\delta})
\nonumber \\
& + & {J \over 2} \sum_{i \delta \sigma} ({\bf S}_i \cdot {\bf S}_{i + \delta}
- {1
\over 4} n_i n_j)~.
\end{eqnarray}
\(H\) acts on a square lattice on which double occupancy is forbidden
($c^{\dag}_{i \sigma} n_i = 0$).  The vectors $\delta(\gamma)$
connect a site with each of its four nearest neighbors (next-nearest
neighbors). While the sign of $t$ can be absorbed in a change of
phase of half the creation operators, the signs of $t'$ and $t''$
are relevant and depend on the form chosen for the restricted
Hilbert space and the precise meaning of $c^{\dag}_{i \sigma}$.
Here this operator creates a {\it particle} at the vacant site $i$
\cite{note}.
\par A recent systematic analytical study of the mapping of the
three-band Hubbard model to $H$ shows that for O-O hopping $t_{pp}
{}~_>^{\sim}$ 0.3eV, the mapping using localized non-orthogonal Zhang-Rice
singlets is more accurate than the one which uses orthogonal O
Wannier functions \cite{ali}.  Using the former mapping, $t_{pp}$
gives rise to a positive contribution to $t''$ due to the
non-orthogonality of the states.  As shown in Fig.~1, numerical
fitting of the parameters of $H$ agrees very well with the
analytical calculations and both show that $t'$ and $t''$ change
sign for increasing $t_{pp}$.
Constrained density-functional calculations predict
$t_{pp}$=0.65eV \cite{hyb} and this results in positive values of
$t'$ and $t''$.  A positive $t'$ \cite{note} agrees with
calculations of other authors \cite{hyb,japo}, with the form of the
Fermi surface \cite{japo}, and with calculations of different
normal state properties \cite{chi}.  Also, the fact that $t' >$ 0
for hole-doped systems while $t' <$ 0 in electron doped systems
\cite{hyb,ali,japo} allows to understand the different relative
stability of the Neel order upon doping in both types of systems
\cite{ali,japo}.  Thus, the parameters taken here: $J = 0.4, t'
\sim 0.2$, and $t'' \sim 0.1$ in units of $t=1$, are close to those
which best describe the hole-doped cuprates.
\par The effect expected from the correlated hopping $t''$ becomes
clear writing the corresponding term in the form of a hopping of
nearest-neighbor singlets:
\begin{equation}
H_{t''} = 2t'' \sum_{i \delta \neq \delta'} b^{\dag}_{i \delta '}
b_{i \delta}~;~~~ b^{\dag}_{i \delta} = {1 \over \sqrt{2}}
(c^{\dag}_{i + \delta \uparrow} c^{\dag}_{i \downarrow} -
c^{\dag}_{i + \delta \downarrow} c^\dagger_{i \uparrow})~.
\end{equation}
Thus, $H_{t''}$ is expected to favor a Bose condensate of singlets
with $d_{x^2-y^2}$ ($s$) symmetry for $t'' > 0 (t'' < 0)$, similar
to the RVB state constructed by Anderson \cite{and}.  This agrees
with mean-field calculations \cite{kot}, and the above mentioned
results for the Hubbard model \cite{chen} (for which $t'' = -J/4$)
and the three-band Hubbard model with $t_{pp} = 0$ \cite{dopf}
(implying also $t'' < 0$ in the effective model $H$).  Instead $J$
at most tends to stabilize static singlets, while $t$ and part of
the terms proportional to $t'$, break the singlets.  Since $d$-wave
pairing is expected to arise from antiferromagnetic spin
fluctuations at small doping \cite{dag,sev}, one expects that a
positive $t''$ would interfere constructively with any
superconducting mechanism already present in $H-H_{t''}$.
\par We have used the Lanczos method to obtain the ground state of
$H$ for a $4 \times 4$ cluster with periodic boundary conditions in
each subspace defined by the $z$ projection of the total spin and
the irreducible representation of the space group.  In Fig.~2 we
show the susceptibility $\chi^d_{sup} = (\sum_{i < j \delta
\delta'} f(\delta) f(\delta ') < g \mid b^\dagger_{i \delta} b_{j \delta
'} \mid g>) /L$ where $\mid g>$ is the ground state, $L$ the size of the
system, $f(\delta) =1$ if
$\delta \parallel {\bf \hat{x}}$ and $f(\delta) = -1$ if $\delta
\parallel {\bf \hat{y}}$ \cite{dag} as a function of $t''$.  The
abrupt increase in $\chi^d_{sup}$ is apparent.  After this
increase, $\chi^d_{sup}$ reaches values similar or larger than the
maximum one obtained previously in the $t-J$ model $(\chi^d_{sup}
\sim 2.5$ for $J=3, x = 0.5, t' = t'' = 0)$ \cite{dag}.  Also, the
qualitative behavior of the superconducting correlation functions
(shown below) changes and they saturate rapidly with distance for $x=0.25$
but not for $x \geq 0.5$. As will be described in detail elsewhere,
except for very particular values of the parameters,
$\chi^d_{sup}$ has its maximum value at a doping
level $x \sim 0.25$ in agreement with experiment.  For $x = 0.75,
\chi^d_{sup} \sim 10^{-2}$ for $t'=0.2$.  For all parameters
considered here the system is far from the region of phase
separation \cite{dag}, which is reduced by $t''$.
\par Our results suggest that as $t''$ is increased, there is a
continuous, although abrupt transition from a disordered spin
liquid to a RVB state with off-diagonal long range order (ODLRO).
To support this conclusion we consider the following simple
generalization to $ x \neq 0$ and ODLRO of Sutherland's RVB wave
function \cite{read}:
\begin{equation}
\mid RVB0 > = P_N \prod_{j \varepsilon A} (1 + \sum_{\delta} g
(\delta) b^\dagger_{i \delta}) \mid 0 >,
\end{equation}
where $P_N$ is the projector over a definite even number of particles $N$,
$j \varepsilon A$ indicates that the product over sites is
restricted to one sublattice $g(\delta) = 1$ for each ``horizontal''
singlet $(\delta \parallel {\bf \hat{x}})$ and $g(\delta) = i$ for
each ``vertical '' singlet $(\delta \parallel {\bf \hat{y}})$.
Eq.~(3) is clearly not an optimum wave function, since unlike
Anderson's one \cite{and} it contains no wave-vector dependence.
In spite of this we find that from its real part and the first few
powers of $H$ applied to it, an excellent variational function for
$t'' > 0.12$ can be constructed.
The real (imaginary) part of $\mid
RVB0>$ has an even (odd) number of vertical singlets.
For $N=L$, $Im \mid RVB0>$=0. It is easy
to check that if the number of singlets $N/2$ is a multiple of four,
$Re \mid RVB0>$ transforms like the representation $\Gamma_1$
{}~(invariant) of the space group, while $Im \mid RVB0>$
transforms like $\Gamma_3~(x^2 - y^2)$.  Instead for
$N/2$ even but not multiple of four, the real (imaginary) part of $\mid RVB0>$
transforms like $\Gamma_3~(\Gamma_1)$ under the symmetry operations of the
space group.
In the $4\times 4$ cluster we find that for $N = 12~ (x =0.25)$ and $N=8~ (x =
0.5)$ for $t'' \geq 0.1$, the ground state $\mid g>$ belongs to the same
irreducible
representation as $Re \mid RVB0>$.  The simplest function which
contains configurations with even and odd number of vertical bonds
with the correct symmetry, and takes advantage of $H_{t''}$ has the form:
\begin{equation}
\mid RVB1> = F (1 - {\alpha \over t''} H_{t''}) Re \mid RVB0>~,
\end{equation}
where $F$ is a normalization factor.  We have determined $\alpha$
maximizing the square of the overlap $S^2 = \mid <g \mid RVB1>
\mid^2$.  This quantity as a function of $t''$ is shown in Fig.~3.
Taking into account the simplicity of Eq.~(4), that there is only
one free parameter, and that the size of the Hilbert space is
huge, the magnitude of $S^2$ after the transition is noticeable.
\par $\mid RVB1>$ can be improved considerably if a variational
function is constructed with $H^n \mid RVB1>~(n=0$~to 3).
The generalized RVB
state thus obtained correctly takes into account
the effects  of $t$ and $t'$ and practically coincides with $\mid
g>$.  However, we will show that $\mid RVB1>$ reproduces already,
at least qualitatively, the main numerical results.  $\chi^d_{sup}$
obtained from $\mid RVB1>$ is compared in Fig.~2.  Note that the
expectation value of $b^{\dag}_{i \delta} b_{i \delta'}$ in $\mid
RVB1>$ for $\delta \perp \delta '$ vanishes for $\alpha = 0$,
implying equal amounts of $s-$ and $d$-wave superconductivity.  For
$t'' > 0.12$ the $d$-wave component of $\mid RVB1>$ dominates but
there is always some $s$ component.
\par In Fig.~4 we
show how after the transition with increasing $t''$, a peak in
($\pi, \pi$) develops in the magnetic structure factor, which is in
excellent agreement with the result of $\mid RVB1>$.  This peak is
probably related with the strong peak in the staggered spin
susceptibility $\chi (\pi,\pi)$ assumed in the phenomenological
spin-fluctuation theories \cite{sev}. For $t''=0$, $S(\pi,\pi)$ slightly
decreases with increasing $t'$. The absence of a structure in $S(\pi/2,\pi/2)$
for $\mid RVB1>$, is due to the lack of magnetic correlations beyond nearest
neighbors in this function, and is restored if $\mid RVB1>$ is improved
including the effect of $t$.
\par In Fig.~5 we show the distance dependence of the $d$-wave
superconducting correlation functions $c(m) = \sum_{i \delta
\delta'} f(\delta) f(\delta ') < g \mid b^\dagger_{i \delta} b_{i + m
\delta '} \mid g>$ and the corresponding expectation value in $\mid
RVB1>$.  One can see that  $c(m)$ saturates at large distances,
suggesting ODLRO for the infinite system.  The significant overlap
of $\mid g>$ with a function which has ODLRO in the thermodynamic
limit is another important indication.  Instead, the corresponding
$s$-wave correlation functions decay steadily with distance.
\par In summary, we have found strong evidence of $d$-wave
superconductivity in the model Eq.~(1) for reasonable $t'' > 0.12$
and experimentally relevant doping $x = 0.25$, and parameters $J=
0.4, t' \sim 0.2$ in units of $t$.  The possibility
of some amount of $s$-wave superconductivity remains open.  The
model and parameters are close to the most realistic ones for the
cuprates.  The superconductivity is rather insensitive to $t'$ and
preliminary results suggest that it persists if a repulsion between
nearest-neighbor Zhang-Rice singlets smaller than $J$ is included
in the model.  The ground state of the system for the above
mentioned parameters is a generalized superconducting RVB state
\cite{and}. A similar
conclusion has been recently obtained for single-rung $t-J$ ladders \cite{sig}.
As noted by Anderson, the RVB state is favored by
electron-phonon interaction, particularly if the system is near an
instability against a Cu dimerization mode \cite{and}.
Theoretical calculations have suggested that this might well be the
case \cite{bar,ali2}.  Also a RVB state allows for a qualitative explanation
of the phase diagram of the high-Tc systems at mean-field level
\cite{lee}, and is consistent with a strongly enhanced $\chi (\pi,
\pi)$ \cite{sev} and with the spin gap observed in YBaCuO
\cite{mig}.  To our knowledge this is the first time that
large superconducting correlations at distances of a few lattice parameters and
the physics of the  RVB is unambiguously obtained numerically in the
ground state of a model for hole-doped CuO$_2$ planes for parameters close
to the optimum ones.
\par We would like to thank M. Bali\~{n}a for providing us with
useful numerical subroutines he developed, and to M.E. Sim\'on, M.
Bali\~{n}a, E. Gagliano, B. Alascio and E. Dagotto for useful
discussions.  One of us (CDB) is supported by the Consejo Nacional
de Investigaciones Cient\'{\i}ficas y T\'{e}cnicas (CONICET), Argentina.
(AAA) is partially supported by CONICET.

\newpage
\section*{Figure Captions}
\noindent{\bf Fig.1:} Parameters of $H$ determined from the three-band Hubbard
model as a function of the O-O hopping $t_{pp}$.  The remaining
parameters were taken from Ref.~\cite{hyb} and the Cu-O hopping
$t_{pd}$ was taken as the unit of energy.  Full line: numerical fit
of the energy levels of a Cu$_4$O$_8$ cluster.  Dashed line:
analytical results from a mapping using localized non-orthogonal
Zhang-Rice singlets.\\
\noindent{\bf Fig.2:} $d$-wave susceptibility as a function of $t''$ for $t=1,
J = 0.4$
and doping $x=0.25$.
Full line: $t'=0.2$ . Dashed line: $t' = 0$. Dotted line: result
using $\mid RVB1>$ (Eq.~(4)) with $t' = 0.2$.\\
\noindent{\bf Fig. 3:} Square of the overlap between the ground state and
$\mid RVB1>$ as a function of $t"$ for $t' = 0.2$. Other parameters
as in Fig. 2.\\
\noindent{\bf Fig. 4:}
Magnetic structure factor $\sum_{ij} <g \mid S^z_i S^z_j
e^{i {\bf q}({\bf R}_i - {\bf R}_j)}\mid g> / L$ as a function of wave vector.
Dashed line: $t'= t'' = 0$.  Full line: $t' = t'' = 0.2$.  Dotted line: result
using $\mid RVB1>$ with $t' = t'' = 0.2$. Other parameters as in Fig. 2.\\
\noindent{\bf Fig. 5:}
$d$-wave superconducting correlation functions as
functions of the distance.  Parameters are the same as in Fig.~4.\\

\begin{references}
\bibitem{dag}E. Dagotto and J. Riera, Phys. Rev. Lett. {\bf 70}, 682 (1993).
\bibitem{chen}G.J. Chen, R. Joynt, F.C. Zhang, and C. Gros, Phys.
Rev. B {\bf 42}, 2662 (1990); Q.P. Li, B.E.C. Koltenbah, and R. Joynt,
Phys. Rev. B {\bf 48}, 437 (1993).
\bibitem{eme}V.J. Emery, Phys. Rev. Lett. {\bf 58}, 2794 (1987).
\bibitem{lit}P.B. Littlewood, C.M. Varma, and E. Abrahams, Phys.
Rev. Lett. {\bf 63}, 2602 (1989) and references therein.
\bibitem{dopf}G. Dopf, A. Muramatsu, and W. Hanke, Phys. Rev. Lett.
{\bf 68}, 353 (1992).
\bibitem{and}P.W. Anderson, Science {\bf 235}, 1196 (1987).
\bibitem{zr}F.C. Zhang and T.M. Rice, Phys. Rev. B {\bf 37}, 3759 (1988).
\bibitem{good}R.J. Gooding and V. Elser, Phys. Rev. B {\bf 41},
2557 (1990);  F.C. Zhang and T.M. Rice, {\it ibid} 2560.
\bibitem{hyb}M.S. Hybertsen, E.B. Stechel, M. Schl\"{u}ter, and
D.R. Jennison, Phys. Rev. B {\bf 41}, 11068 (1990).
\bibitem{ding}H.Q. Ding, G.H. Lang, and W.A. Goodard III, Phys.
Rev. B {\bf 46}, 14317 (1992).
\bibitem{bat}C.D. Batista and A.A. Aligia, Phys. Rev. B {\bf 48}, 4212
(1993); {\bf 49}, 6436 (E) (1994).
\bibitem{ali}A.A. Aligia, M.E. Sim\'on, and C.D. Batista, Phys.
Rev. B. 49, 13061 (1994)
\bibitem{chi}H. Chi and A.D.S. Nagi, Phys. Rev. B {\bf 46}, 421
(1992); M.J. Lercher and J.M. Wheatly, Physica C {\bf 215}, 145
(1993); R.J. Gooding, K.J.E. Vos, and P.W. Leung, Phys. Rev. B {\bf
49}, 4119 (1994).
\bibitem{japo}T.Tohyama and S. Maekawa, Phys. Rev. B {\bf 49}, 3596 (1994).
\bibitem{note}If the Zhang-Rice singlets were represented by doubly
occupied sites as in Ref.~\cite{hyb}, or if the $c^{\dag}_{i
\sigma}$ would destroy particles creating vacant sites, all $t$'s
would change sign.  In most studies of $t-J$- like models the
particles are called ``electrons'' and the vacant sites ``holes'',
although for hole-doped systems they represent respectively holes
and Zhang-Rice singlets of the original multiband Hubbard model.
\bibitem{kot}G. Kotliar, Phys. Rev. B {\bf 37}, 3664 (1988).
\bibitem{sev}N.E. Bickers D.J. Scalapino, and S.R. White, Phys.
Rev. Lett. {\bf 62}, 961 (1989); P. Monthoux, A.V. Balatsky, and D.
Pines, Phys. Rev. Lett. {\bf 67}, 3448 (1991); P. Monthoux and D.
Pines, Phys. Rev. Lett. {\bf 69}, 961 (1992).
\bibitem{read}N. Read and B. Chakraborty, Phys. Rev. B. {\bf 40},
7133 (1989); and references therein.
\bibitem{sig} S. Gopalan, T. M. Rice and M. Sigrist,
Phys. Rev. B {\bf 49}, 8901 (1988);
M. Sigrist, T. M. Rice and F. C. Zhang, Phys. Rev. B
(to be published).
\bibitem{bar}S. Barisi\v{c}, I. Batisti\v{c}, and J. Friedel, Europhys. Lett.
{\bf
3}, 1231 (1987).
\bibitem{ali2}A.A. Aligia, Phys. Rev. B {\bf 39}, 6700 (1989).
\bibitem{lee}P.A. Lee and N. Nagaosa, Phys. Rev. B {\bf 46}, 5621 (1992).
\bibitem{mig}J. Rossat-Mignod {\it et al}, Physica C {\bf 185-189}, 86
(1991); Physica B {\bf 186-188}, 1 (1993).
\end{references}
\end{document}